\documentclass[jgrga]{agu2001}
\ifgalley
\let\eject\relax
\fi
\usepackage{agu-ps}
\usepackage{epsfig}
\usepackage{graphicx}

\lefthead{Gangopadhyay et al.}

\righthead{Interpretation of Pioneer 10 Ly- $\alpha$}

\journalid{Sometime 2002} \articleid{???}{??}
\paperid{2002???00021} \cpright{AGU}{2002} 

\received{February 23, 2002} \revised{} \accepted{1} \published{}

\authoraddr{P Gangopadhyay,University of Southern California, Los Angeles
 (pradip@lism.usc.edu)}

\authoraddr{Vlad Izmodenov,
Lomonosov Moscow State University, Department of Aeromechanics and
Gas Dynamics, Faculty of Mechanics and Mathematics, Moscow,
119899, Russia (izmod@ipmnet.ru)}

\begin{document}

\title{Interpretation of Pioneer 10 Ly-
$\alpha$ based on heliospheric interface models: methodology and
first results}

\author{Pradip Gangopadhyay$^{1}$, Vlad Izmodenov$^{2}$, Mike Gruntman$^{1}$, Darrell Judge$^{1}$}
\affil{(1) University of Southern California, Los Angeles} \affil{
(2) Lomonosov Moscow State University, Department of Aeromechanics
and Gas Dynamics, Faculty of Mathematics and Mechanics, Moscow,
119899, Russia}

\begin{abstract}
The Very Local Interstellar Medium (VLISM) neutral hydrogen and
proton densities are still not precisely known even after three
decades of deep space research and the existence of the EUV and
other diagnostic data obtained by Pioneer 10/11, Voyager 1/2 and
other spacecraft. The EUV data interpretation, in particular,  has
suffered because of inadequate neutral hydrogen-plasma models,
difficulty of calculating the multiply scattered Lyman $\alpha$
glow and calibration uncertainties. Recently, all these
difficulties have been significantly reduced. In the present work
we have used the latest state of the art supersonic VLISM neutral
hydrogen-plasma and Monte Carlo radiative transfer model,
incorporating neutral density, temperature, and velocity
variations, actual solar line shape, realistic redistribution
function, Doppler and aberration effects.  This work presents the
methodology of the radiative transfer code and the first results
of the comparison of the model predictions with the Pioneer 10
data. Monte Carlo radiative transfer calculations were carried out
for five neutral hydrogen- plasma models and compared with Pioneer
data. The first results are quite encouraging. We found that the
VLISM ionization ratio is between 0.2 and 0.5  and that the VLISM
neutral hydrogen density is less than 0.25 cm$^{-3}$. The present
calculation suggests that the Pioneer 10 photometer derived
intensities (Rayleighs) need to be increased by a factor of 2. If
this model- derived calibration is used then the difference
between Pioneer 10 and Voyager 2 intensity values is reduced to
about 2.2. The model, neutral hydrogen density=0.15 cm$^{-3}$ and
proton density=0.07 cm$^{-3}$, is found to best fit the Pioneer 10
data.
\end{abstract}

\begin{article}
\section{Introduction}

The heliospheric interface, formed due to the interaction between
the solar wind and the local interstellar cloud (LIC), is a very
complicated phenomenon where the solar wind and interstellar
plasmas, interstellar neutrals, magnetic field, and cosmic rays
play prominent roles [{\it Axford}, 1972; {\it Holzer}, 1972]. The
heliosphere provides a unique opportunity to study in detail the
only accessible example of a commonplace but fundamental
astrophysical phenomenon - the formation of an astrosphere. The
heliospheric interface is a natural "environment" of our star and
knowledge of its characteristics is important for the
interpretation and planning of space experiments. In fact, the
growing body of evidence about the interface using different sets
of data including Lyman $\alpha$ data [{\it Hall et al.},1993;
{\it Kurth and Gurnett}, 1993; {\it Cummings, Stone and Webber},
1993; {\it Linsky and Wood}, 1996; {\it Gloeckler and Geiss},
2001; {\it Wang and Richardson}, 2001] shows that the properties
of the neutral hydrogen distribution are significantly affected by
processes in the interface between the interstellar medium and the
solar wind and that a 'no interface model' can not explain the
observations. Thus it is specially important to include the effect
of the interface in the interpretation of deep space spacecraft
data. In this work we have included the effect of the interface in
the interpretation of the Pioneer 10 photometer hydrogen channel
data.
\begin{figure} \label{interf_struct}

\noindent\includegraphics[width=10cm,clip]{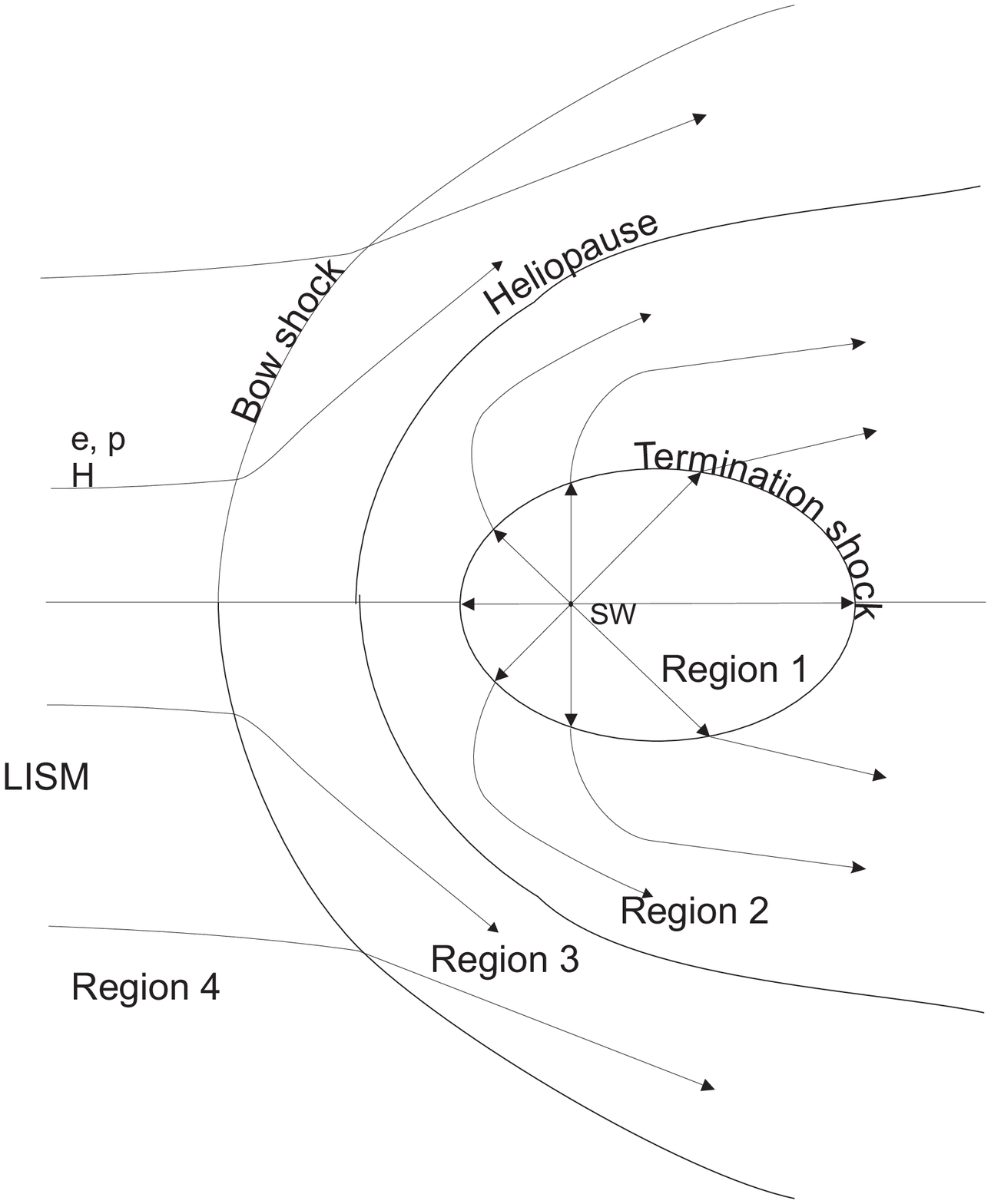}
\caption{The heliospheric interface is the region of the solar
wind interaction with LIC. The heliopause is a contact
discontinuity, which separates the plasma wind from interstellar
plasmas. The termination shock decelerates the supersonic solar
wind. The bow shock may also exist in the interstellar medium. The
heliospheric interface can be divided into four regions with
significantly different plasma properties: 1) supersonic solar
wind; 2) subsonic solar wind in the region between the heliopause
and termination shock; 3) disturbed interstellar plasma region (or
"pile-up" region) around the heliopause; 4) undisturbed
interstellar medium. }
\end{figure}

Remote sensing of the heliospheric interface through the study of
the interstellar hydrogen atoms is possible since H atoms play a
very important role in the formation of the heliospheric
interface. Interstellar H atoms are strongly coupled with plasma
protons by charge exchange. The charge exchange process leads to a
significantly smaller heliosphere.
 The interstellar neutral hydrogen atoms penetrate deeply
into the heliosphere since the mean free path is comparable to the
size of the heliosphere. Inside the heliosphere, the atoms and
their derivatives such as pickup ions and anomalous cosmic rays
are measured. These species become major sources of observational
information on the heliospheric interface and therefore on the
local interstellar medium properties. For example, direct
detection of the flux of the interstellar neutral helium allows
the determination of the Sun/LIC relative velocity and local
interstellar temperature [{\it Witte et al.}, 1993, 1996].
Interstellar helium atoms are not affected by the heliospheric
interface plasma due to their small charge exchange cross section
with protons. In contrast to helium, interstellar hydrogen atoms
have strong coupling with heliospheric plasma protons.
Distribution of these atoms inside the heliosphere has imprints of
the heliospheric interface.  Thus, interstellar hydrogen atoms
provide excellent remote diagnostics on the structure of the
heliospheric interface.

The study of the neutral hydrogen atoms in the outer heliosphere
has been made possible by the presence of four deep space
spacecraft, Pioneers 10 and 11 (P10 and P11), and Voyagers 1 and 2
(V1 and V2).  The USC photometers on-board P10 and P11 and the
ultraviolet spectrometers (UVS) on-board V1 and V2 have measured
the interplanetary Ly $\alpha$ background radiation for more than
twenty years. Various studies of P10/11 and V1/2 Ly $\alpha$ data
have been published [{\it Wu et. al.}, 1981, 1988; {\it Shemansky
et al.}, 1984; {\it Ajello et al.}, 1987; {\it Gangopadhyay et
al.}, 1989; {\it Hall et al.}, 1993;
 Qu\'emerais et al., 1995,1996; Gangopadhyay and Judge, 1995, 1996]. Yet, the
estimation of the interstellar H atom density varies greatly from
study to study, ranging between 0.03 and 0.3 cm-3 [see, {\it
Quemerais et al.}, 1994].

{\it Hall et al.} [1993] found that the Ly $\alpha$ intensity
falls with heliocentric distance less quickly than expected from a
standard hot model. This result suggested that there was a
positive gradient of H atom density at large distances from the
Sun. This can be explained by a hydrogen wall around the
heliosphere confirmed by {\it Lallement et al.} [1993]. However,
{\it Quemerais et al.} [1995] suggested an alternative
explanation. The latter work suggested that the increase of Ly
$\alpha$ intensity in the upwind direction could be partially due
to the constant emission from HII regions in the galactic plane.

The study of the heliosphere has been made more difficult by the
calibration difference of 4.4 at Lyman $\alpha$ between the V1/2
spectrometers and P10/11 photometers found by {\it Shemansky et
al.} [1984]. This difference was determined by comparing V2 and
P10 data when both spacecraft looked in the same direction and at
the same time. Finally, the picture is further complicated by the
change in the absolute value of the solar Ly $\alpha$ flux since
1993. Recently, past solar Lyman $\alpha$ irradiance measurements
have been consolidated into a long-term composite time series
[Woods and Rottman, 1997; Tobiska et al. 1997; Woods et al.,
2000]. This was done by adjusting Atmospheric Explorer E (AE-E)
and the Solar Mesospheric Explorer (SME) to agree with the Upper
Atmospheric Research Satellite (UARS) Solar Stellar Irradiance
Comparison Experiment (SOLSTICE) data. This resulted in changes in
reported solar Lyman $\alpha$ measurements obtained by both the
SME and the AE-E satellites. Of course, the calculation reported
here assumes that the line center flux varies the same way as the
total flux. If this assumption does not hold then that will
introduce an additional uncertainty.

In this paper we will describe the methodology and the first
results of our reanalysis of the P10 Ly $\alpha$ data to improve
our knowledge of the very local interstellar neutral hydrogen and
proton densities. This reanalysis uses the latest state of the art
neutral hydrogen-plasma and radiative transfer models outlined in
the later sections.

\begin{figure} \label{fig_Hatoms}
\noindent\includegraphics[width=\hsize]{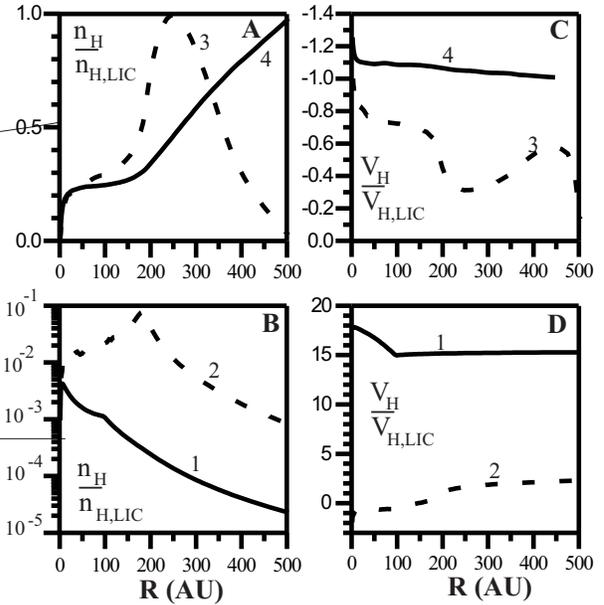}
\caption{Number densities and velocities of four populations of H
atoms as functions of heliocentric distance in the upwind
direction. 1 designates atoms created in the supersonic solar
wind, 2 atoms created in the heliosheath (SSWAs), 3  atoms created
in the disturbed interstellar plasma (HIAs), and 4 original (or
primary) interstellar atoms (PIAs). Number densities are
normalized to $n_{H,LIC}$, velocities are normalized to $V_{LIC}$.
}
\end{figure}

\begin{figure} \label{fig3}
\noindent\includegraphics[width=\hsize]{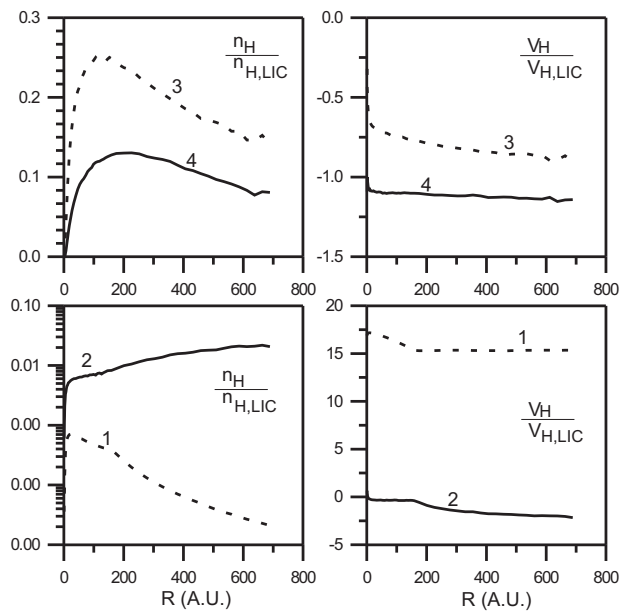}
\caption{Number densities and velocities of four populations of H
atoms as functions of heliocentric distance in the downwind
direction. Notations are the same as in figure 2.}
\end{figure}

\section{Model of the H atom distribution}

The interaction of the solar wind with the interstellar medium
influences the distribution of interstellar atoms inside the
heliosphere. Further, it is now clear that the Local Interstellar
Cloud is partly ionized and that the plasma component of the LIC
interacts with the solar wind plasma to form the heliospheric
interface (Figure 1). Interstellar H atoms interact with the
plasma component through charge exchange. This interaction
strongly influences both the plasma and neutral components. The
main difficulty in the modeling of the H atom flow in the
heliospheric interface is its kinetic character due to the large,
i.e. comparable to the size of the interface, mean free path of H
atoms with respect to the mean free path for charge exchange
process. Several models have been developed [e.g., {\it Baranov
and Malama}, 1993, 1995, 1996; {\it M\"{u}ller et al.}, 2000]. In
this paper to get the H atom distribution in the heliosphere and
heliospheric interface structure, we use the self-consistent model
developed by Baranov and Malama [1993]. The kinetic equation for
the neutral component and the hydrodynamic Euler equations were
solved self-consistently by the method of global interactions. To
solve the kinetic equation for H atoms, an advanced Monte Carlo
method with splitting of trajectories [{\it Malama}, 1991] was
used. Basic results of the model were reported by {\it Baranov and
Malama} [1995], {\it Izmodenov et al.} [1999], {\it Izmodenov}
[2000], and {\it Izmodenov et al.} [2001].

Hydrogen atoms newly created by charge exchange have the
velocities of their ion partners in charge exchange collisions.
Therefore, the parameters of these new atoms depend on the local
plasma properties. It is convenient to distinguish four different
populations of atoms depending on where in the heliospheric
interface they originated. Population 1 corresponds to the atoms
created in the supersonic solar wind noted as {\bf SSWA}
(supersonic solar wind atoms). Population 2 (noted {\bf HSWA}, hot
solar wind atoms) represents the atoms originating in the
heliosheath and known as heliospheric ENAs [{\it Gruntman et al.},
2001]. Population 3 ({\bf HIA}, hot interstellar atoms) consists
of the atoms created in the disturbed interstellar wind. We will
call original (or primary) interstellar atoms population 4 ({\bf
PIA}, primary interstellar atoms). The number densities and mean
velocities of these populations are shown  in Figures 2 and 3 as
functions of the heliocentric distance in the upwind and downwind
directions respectively.

The main results of the model for the H atom populations can be
summarized as follows:

{\bf PIAs} are significantly filtered (i.e. their number density
is reduced) before reaching the termination shock. Since slow
atoms have a smaller mean free path as compared with fast atoms,
they undergo more charge exchange. This kinetic effect, called
{\em selection}, results in a deviation of the interstellar
distribution function from Maxwellian [{\it Izmodenov et al.},
2001]. The selection also results in $\sim$10 \% increase of the
primary atom mean velocity at the termination shock (Figure 2C).

{\bf HIAs} are created in the disturbed interstellar medium by
charge exchange of primary interstellar neutrals and protons
decelerated by the bow shock. The secondary interstellar atoms
collectively make up the {\em H wall}, a density increase at the
heliopause. The {H wall} has been predicted by {\it Baranov et
al.} [1991] and confirmed by various observations [e.g. {\it
Lallement et al.}, 1993; {\it Linsky and Wood}, 1996; {\it
Gloeckler et al.}, 1997]. At the termination shock, the number
density of the secondary neutrals is comparable to the number
density of the primary interstellar atoms (Figure 2A, dashed
curve). The relative abundances of PIAs and HIAs entering the
heliosphere depends on the degree of ionization in the
interstellar medium. It has been shown by {\it Izmodenov et al.}
[1999] that the relative abundance of HIAs inside the termination
shock increases with increasing interstellar proton number
density. The bulk velocity of HIAs is about 18-19 km/s. This
population approaches the Sun. The velocity distribution of HIAs
is not Maxwellian. The velocity distributions of different
populations of H atoms were calculated in {\it Izmodenov et al.}
[2001] for different directions from upwind. The fine structures
of the velocity distribution of the primary and secondary
interstellar populations vary with direction. These variations of
the velocity distributions reflect the geometrical pattern of the
heliospheric interface. The velocity distributions of the
interstellar atoms can provide good diagnostics of the global
structure of the heliospheric interface.

The third component of the heliospheric neutrals, {\bf HSWAs},
corresponds to {\bf the neutrals created in the heliosheath} from
hot and compressed solar wind protons. The number density of this
population is an order of magnitude smaller than the number
densities of the primary and secondary interstellar atoms. This
population is of minor importance for interpretations of Ly
$\alpha$ and pickup ion measurements inside the heliosphere.
However, some of these atoms may probably be detected by
Ly$\alpha$ hydrogen cell experiments due to their large Doppler
shifts. Due to their high energies, the particles influence the
plasma distributions in the LIC. Inside the termination shock the
atoms propagate freely. Thus, these atoms can be the source of
information on the plasma properties in the place of their birth,
i.e. the heliosheath.

The last population of heliospheric atoms is SSWAs, {\bf the atoms
created in the supersonic solar wind}. The number density of this
atom population has a maximum at $\sim$5 AU from the sun. At this
distance, the number density of population 1 is about two orders
of magnitude smaller than the number density of the interstellar
atoms. Outside the termination shock the density decreases faster
than $1/r^2$ where $r$ is the heliocentric distance (curve 1,
Figure 2B). The mean velocity of population 1 is about 450 km/sec,
which corresponds to the bulk velocity of the supersonic solar
wind. The {\em supersonic} atom population results in the plasma
heating and deceleration upstream of the bow shock. This leads to
the decrease of the Mach number ahead of the bow shock.

SSWAs velocities are Doppler shifted out of the solar H Lyman
$\alpha$ line and therefore are not detectable by interplanetary
Lyman $\alpha$ measurements. Atoms of the three other populations
penetrate the heliosphere and may backscatter solar Ly$\alpha$
photons. This is in contrast to the classical hot model [{\it
Thomas}, 1978; {\it Lallement et al.}, 1985] which assumed that at
large heliocentric distance the velocity distribution is
Maxwellian and unperturbed by the heliospheric interface
interaction.

In what follows, we will use these three populations to compute
the interplanetary UV background. Each population will be referred
to by use of its label (PIA, HIA, HSWA). The hydrogen distribution
model will be called the three population model, noted $3p$ model,
because the SSWA is invisible to Ly$\alpha$ light.

\begin{table*}
\begin{center}
\caption{The P10 data, position and solar flux}
\begin{tabular*}{\hsize}{@{\extracolsep{\fill}}ccccccc}
\tableline
Year & Day  & heliocentric & ecliptic & ecliptic  & Solar      & counts \\
     &      & distance     & latitude &longitude  &Lyman $\alpha$ & per    \\
     &      & (AU)         & of P10   & of P10    & flux       &  sec   \\
     &      &              & measured & measured  &(photons/   & \\
     &      &              & from     & from      &  cm2/s)    & \\
     &      &              & Earth    & Earth     &            & \\
 1979&  298 & 20.0043      &3.257     &59.982     & 5.7e11     & 405.30\\
 1980&  252&  22.5058      &3.325     &62.662     &5.4e11      &
 362.66\\
 1981&  207&  25.0001      &3.347   &64.615      &5.19e11      &
 323.33\\
 1982&  167&  27.5037      &2.561   &65.600      &5.05e11      &
 268.33\\
 1983&  129&  30.0002      &2.589   &65.772      &5.10e11      &
 214.90\\
 1984&   95&  32.5061      &3.308   &66.293      &4.71e11      &
 185.57\\
 1985&   61&  35.0067      &3.030   &67.902      &3.89e11      &
 149.01\\
 1986&   30&  37.5069      &2.380   &68.560      &3.52e11      &
 128.23\\
 1987&    1&  40.0006      &2.842   &69.712      &3.72e11      &
 135.35\\
 1987&  338&  42.5012      &2.628   &71.280      &4.01e11      &
 145.84\\
 1988&  314&  45.0111      &2.544   &72.150      &5.03e11      &
 169.28\\
\tableline
\end{tabular*}
\end{center}
\end{table*}

\begin{table*}
\caption{Sets of model parameters and results}
\begin{tabular*}{\hsize}{@{\extracolsep{\fill}}lcccc}
\tableline
Model &  $n_{H,LIC}$ & $n_{p,LIC}$ & sqrt(Least squares sum) & Calibration factor  \\
 \tableline \\
1 & 0.15 & 0.07 & 40.9 & 2.04   \\
2 & 0.2  & 0.05 & 71.8 & 3.16   \\
3 & 0.2  & 0.10 & 50.7 & 2.26   \\
4 & 0.2  & 0.20 & 50.4 & 1.80   \\
5 & 0.25 & 0.10 & 56.9 & 3.08  \\
\tableline \\
\end{tabular*}
\end{table*}

\section{Radiative transfer model}

The LISM neutral hydrogen gas is without any doubt an optically
thick medium for solar Lyman $\alpha$ photons at the large
heliocentric distances considered here. This is because the
scattering path length for neutral hydrogen density of 0.1
cm$^{-3}$ will be of the order of 10 to 15 AU. This implies that
the radiative transfer calculation of Lyman $\alpha$ photons at
heliocentric distances greater than ~15 AU must necessarily take
into account multiple scattering. In fact, a full treatment of the
solar Lyman $\alpha$ radiative transfer problem must include the
actual self-reversed solar line shape, multiple scattering, full
angular and frequency redistribution function, Doppler and
aberration effects, heliosphere-wide hydrogen temperature and
velocity changes and Voigt Lyman $\alpha$ absorption profile.

The Monte Carlo radiative transfer calculation performed here is a
revised version of the code published in {\it Gangopadhyay, Ogawa
and Judge} [1989]. The original 1989 code agreed with {\it Keller
et al.} [1981] for a hot hydrogen model. The 1989 code included a
flat solar line, multiple scattering, complete frequency
redistribution, constant hydrogen temperature and Doppler
absorption profile. The 1989 model has now been completely revised
to incorporate all the requirements listed in the previous
paragraph.

The radiative transfer code is outlined below:

1. The hydrogen Ly $\alpha$ photons are launched from the sun,
which is the origin of the coordinate system.

2. The direction of launch is determined randomly. $\theta$ is the
angle between the launch direction and the upwind direction and
$\phi$ is the azimuth angle around the interstellar flow direction
which is the line of symmetry. The angles $\theta$ and $\phi$ are
determined from the equations $cos(\theta)=2t-1$ and $\phi=2 \pi
t$, respectively, and t is a random number between 0 and 1.

3. The photon frequency is determined by inverting the solar Lyman
$\alpha$ profile [{\it Lemaire et al.}, 1978].

4. The optical depth of the photon is given by $\tau=-ln(t)$. The
new position of the photon is found by inverting the optical
depth.

5. It is checked whether the photon entered the field of view of a
UV detector with a specific look angle.

6. The photon is followed till it crosses a cut-off optical depth
of 75. Then a new photon is followed by going to step 1.

7. As long as the photon optical depth is less than the cut-off
optical depth, the photon is scattered again. Gamma, the
scattering angle is obtained from the relation:
\[
cos(\gamma)= (t-0.5) \frac{24}{11} - (t-0.5)^3 \frac{8}{11}
\]
 The azimuth angle $\phi$ around the incident direction is
given by $\phi=2 \pi t$.

8. The new frequency $\nu_p$ is obtained by inverting the
redistribution function [{\it Mihalas}, 1978]:
\begin{eqnarray}
R(\nu,\nu_p)= \frac{g}{\pi sin \theta} exp[- \frac{1}{2} (x -
x')^2 cosec^2 \frac{\theta}{2}] \times \\
H( a sec \frac{\theta}{2}, \frac{1}{2} (x + x') sec
\frac{\theta}{2} ), \nonumber
\end{eqnarray}
where g is the angular phase function. $x= (\nu -\nu_0)$/(Doppler
width) and $x' = (\nu' - \nu_0)$/(Doppler width) are dimensionless
frequencies of the incident and scattered photon, where $\nu_0$ is
the line center frequency. $\theta$ is the angle between the
incident and scattered direction.
 $H$ is the Voigt function and a is equal to
($\gamma/4\pi$)/(Doppler width) where $\gamma$ is the radiative
damping width of the upper state.  This redistribution function is
more general than other redistribution functions, like complete
frequency redistribution function, commonly in use. The Mihalas
redistribution function is the correct function to handle the
resonant scattering of the strongly self-reversed solar line in
the interplanetary medium since this function simulates nearly
complete redistribution in the line core and becomes nearly
coherent in the line wings [Mihalas, 1978]. Since the photon
frequency and direction are calculated in the stationary sun
centered frame while the hydrogen atoms have a flow velocity
relative to the stationary frame, it is necessary to calculate the
frequency and incident photon direction in the moving atom frame.
The equations used to carry out this transformation for bulk
velocity significantly less than the speed of light $c$ are given
below:
\begin{eqnarray}
\frac{sin \theta' cos \phi'}{\lambda'} = \frac{sin \theta cos \phi
- v_x/c}{\lambda} \nonumber \\
\frac{sin \theta' sin \phi'}{\lambda'} = \frac{sin \theta sin \phi
- v_y/c}{\lambda} \\
\frac{cos \theta'}{\lambda'} = \frac{cos \theta
- v_z/c}{\lambda} \nonumber \nonumber \\
\nu' = \nu (1 - \frac{(\vec{v} \cdot \vec{k})}{c}) \nonumber
\end{eqnarray}

The moving atom frame is the primed frame. $\vec{k}$ is the photon
direction vector and $\lambda$ is the photon wavelength in the
rest frame.

9. A new optical depth and a new position is chosen exactly as in
step 4.

10. The code goes to step 5 and the subsequent steps are repeated.

11. The intensity in Rayleighs is calculated using the following
expression:
\begin{equation}
4 \pi I =\frac{4 \pi N}{A \Omega} \frac{F_e \cdot 4 \pi
r_e^2}{N_{tot}} 10^{-6},
\end{equation}
where $I$ is the specific intensity, $N$ is the number of photons
collected in the collection area $A$, $\Omega$ is the solid angle
of the collection cone, $F_e$ is the integrated solar Lyman
$\alpha$ flux at 1 AU in photons/cm$^2$/s, $r_e$ is 1 AU and
$N_{tot}$ is the number of photons launched. The Pioneer 10
collection geometry, a 1$^o$ conical ring  pointing anti sunwards
and  centered 160$^o$ with respect to the spacecraft spin axis
pointed approximately towards the sun, was not simulated. A 5$^o$
conical ring was used to improve the statistics. In fact, the
spacecraft spin axis is supposed to point towards the Earth.
However, the spin axis does drift away from the Earth and has to
be occasionally corrected. The angle, Earth-spacecraft-spin axis
is not allowed to exceed 0.5 degrees. Of course, for the
heliocentric distances considered here, the spin axis may be
thought of as pointing approximately towards the sun. The solid
angle, $\Omega$, is given by,
$\Omega=2*\pi*(cos(17.5)-cos(22.5))$, where 22.5 and 17.5 are the
semi vertex angles of the outer and inner collection cones. The
photons are collected on a 20 degree wide patch of the collection
sphere centered at the sun of radius equal to the heliocentric
distance [{\it Gangopadhyay, Ogawa and Judge}, 1989]. The error
(rayleighs) is calculated by dividing the intensity (Rayleighs) by
the square root of the number of photons collected.

\section{Pioneer 10 instrumentation and data}

The Ultraviolet photometers on-board Pioneer 10 cover two broad
spectral regions. The long wavelength channel is sensitive to
emissions shortwards of 1400 $A$, which includes the Ly $\alpha$
emission line. The short wavelength channel is sensitive shortward
of 800 A. The details of the UV photometers and their sensitivity
curves are given in {\it Carlson and Judge} [1974].

The trajectory of Pioneer 10 lies nearly in the ecliptic and is in
the downstream direction relative to the interstellar wind
velocity. The details of the trajectory are given in {\it Wu et
al.} [1988]. The detector look angle traces out a conical shell (
apex angle 40 degrees and shell thickness = 1 degree) about the
spacecraft spin axis. The Pioneer 10 photometer suffered a gain
loss at heliocentric distances greater than about 45 AU,
increasing the uncertainty from 5 \% [{\it Wu et al.}, 1988] for
data within about 45 AU of the Sun to uncertainty values still
under review. The reasons for the gain loss are discussed in {\it
Hall et al.} [1993].

We have used daily averaged Pioneer 10 Ly $\alpha$ data obtained
at heliocentric distances between 20 and 45 AU in the present
work. The inner distance limit, 20 AU, was chosen because our
radiative transfer code does not generate good statistics for data
positions too close to the sun. The outer limit was set at 45 AU
because Pioneer 10 EUV data uncertainty increases sharply beyond
45 AU. The P10 data, position and solar flux are given in Table 1.
The look angle can be calculated from the spacecraft position. For
all the eleven data points, the Earth-spacecraft vector looks away
from the galactic center making an angle of about 109 degrees with
respect to the Galactic North pole. The count rates are converted
to Rayleighs by subtracting the dead count (3 counts per sec) and
dividing by 4.9. Later on we will see that the P10 intensity
values obtained by this procedure need to be revised upwards. A
point that needs to be made here is that our data set and the P10
data published by {\it Scherer and Scherer} [2001] are obtained
from the same set of raw data.  Any difference in the two data
sets would be due to different processing of the raw data in the
post 45 AU data set where P10 hydrogen channel suffered a gain
loss. The difference between the data sets is insignificant for
the pre 45 AU region.

\section{Methodology of comparison of theory and observations}

Monte Carlo radiative transfer calculations were carried out for a
number of neutral hydrogen density models (Table 2). The
calculated results, Icalc, were then compared with P10 EUV data
(figs 4,5,6,7 and 8). In order to properly compare the data with
the calculated results, it was necessary to calculate the optimum
P10 instrument calibration factor (CF) for each of the density
models. This step is necessary since it is known that the P10 and
V1/2 instrumental calibrations differ by a factor of 4.4 at Lyman
$\alpha$ [Shemansky et al. 1984]. The difference between the P10
photometer and V2 spectrometer calibration factors forces one to
reproduce the distance dependence of the data rather than rely on
the absolute value of the measured intensity. It should be stated
here that the P10 photometer calibration did not drift with time
during the period 1979-1988 considered here. The degradation of
the P10 Bendix channel multipliers has been studied in the
laboratory [{\it Carlson and Judge}, 1974]. It has been found that
the electron multipliers can deliver about 16 coulombs of charge
without any sign of fatigue. The P10 electron multiplier for the
hydrogen channel is estimated to have delivered at most 4 coulombs
of charge by 1988. The early degradation observed in the hydrogen
channel of the P11 instrument is attributed to damage due to the
hostile environment encountered by P11 during its flyby past
Saturn. The optimum calibration factor for a density model is
calculated by minimizing the least squares sum, LSS, where LSS is
calculated by the following equation
\begin{equation}
LSS= \sum (I_{model}+ bg - CF*I_{P10 data})^2,
\end{equation}
where summation is over the P10 data points and bg is the Lyman
$\alpha$ galactic background.  Both CF and bg were varied to
obtain the minimum LSS. Once the optimum $CF$ and bg are found
then P10 data are multiplied by $CF$ and compared with the
calculated intensity. Both $CF$ and $LSS$ for each of the 5
density models are given in Table 2.

It is clear from Table 2 and the figures that the model with the
VLISM neutral hydrogen density of 0.15 cm$^{-3}$ and proton
density of 0.07 cm$^{-3}$ yields the lowest $LSS$ and so best
reproduces the P10 data. The next best fit occurs for the model
with neutral hydrogen density of 0.2 cm$^{-3}$ and proton density
of 0.2 cm$^{-3}$. It is not at present possible to choose between
the two neutral densities used in these two models as it is
necessary to calculate model results for other ionization ratios
for both of these cases and compare with P10 data. The Lyman
$\alpha$ background, bg, was determined to be negligibly small for
all the density models. In fact the best fit was obtained for bg
equal to zero although the deviation of the data from the best fit
curve (Figure 4) was from + 22 to - 16 Rayleighs. The background
glow is assumed to be approximately the same for all eleven data
points since the look directions are approximately the same with
respect to the galactic plane. A look at Table 2 and the figures
also show that the largest deviation from the P10 data occurs for
the model in which neutral density is 0.2 cm$^{-3}$ and proton
density=0.05 cm$^{-3}$. The ionization ratio (n$_p$/(n$_H$+n$_p$))
for this model is 0.2.

The value 0.2 might well be the lower limit of the LISM ionization
ratio. This is because the LISM neutral hydrogen density can not
be too high because of the growing evidence that the neutral
hydrogen density inside the termination shock is of the order of
0.1 cm$^{-3}$ or less. For example, {\it Wang and Richardson}
[2001] suggest that the Voyager 2 solar wind observations are best
fitted with an interstellar neutral hydrogen density of 0.08
cm$^{-3}$ at the termination shock. Such a low neutral hydrogen
density at the termination shock would imply a neutral hydrogen
density of 0.16 cm$^{-3}$ at "infinity" even assuming that the
neutral hydrogen suffers a 50 \% depletion at the interface.
Similarly, {\it Gloeckler and Geiss} [2001] found that the neutral
hydrogen density at the termination shock is 0.115 cm$^{-3}$ and
obtained a value of 0.18 cm$^{-3}$ for the interstellar hydrogen
density assuming a 58 \% filtration effect. We did not use a very
high interstellar neutral hydrogen density as that will imply a
large neutral density inside the heliosphere which would
contradict observational evidence [{\it Gloeckler and Geiss},
2001; {\it Wang and Richardson}, 2001].

 It is possible to estimate the upper limit to the
ionization ratio from the fact that the model with neutral and
proton densities equal to 0.2 cm$^{-3}$ (ionization ratio =0.5)
gives better fit to the Pioneer 10 data than the models discussed
previously except for the first model in Table 2. However, such a
high proton density is ruled out as it would imply a solar wind
shock too close to the sun contrary to observations. Thus, an
ionization ratio as high as 0.5 would imply a low neutral density
and a ratio higher than 0.5 is extremely unlikely. The relatively
high value of the ionization ratio (0.2 to 0.5) estimated in this
work clearly shows that the VLISM neutral hydrogen density can not
be as high as 0.25 cm$^{-3}$. This is because it is clear from
figure 8 and from the better fit obtained for the model with both
proton and neutral densities equal to 0.2 cm$^{-3}$ that an
ionization ratio substantially higher than 0.3 would be necessary
to better fit the neutral density, 0.25 cm$^{-3}$, model with the
P10 data. However, even an ionization ratio of 0.4 for a neutral
hydrogen density of 0.25 cm$^{-3}$ would imply a proton density of
about 0.18 cm$^{-3}$ which would move the solar wind termination
shock too close to the sun.

Another issue that needs to be discussed is the possible reasons
for the deviation of the model calculation from the P10 data. The
obvious reason is, of course, the difference of the model neutral
hydrogen density from the actual heliospheric neutral density.
Another reason is the possible variation of the solar Lyman
$\alpha$ line center flux with respect to the integrated line
[{\it Lemaire et al.}, 1998].  We have plotted the ratio of the
model intensity to the P10 data against solar Lyman $\alpha$ flux
in order to see if the deviation is due to the variation in line
center flux. There is a trend for the ratio to decline from
greater than 1 to less than 1 as the solar flux increases. This
trend might be due to the line center flux variation.

Figure 4 also shows a comparison between optically thick and
optically thin radiative transfer calculations. The difference is
significant. This is a direct confirmation that the interstellar
gas is optically thick medium and does not support the recent
claims of Scherer and Fahr, 1996, Scherer, Fahr and Clarke, 1997
and Scherer et al. 1997 that the gas is optically thin.

\begin{figure} \label{fig4}
\noindent\includegraphics[width=\hsize]{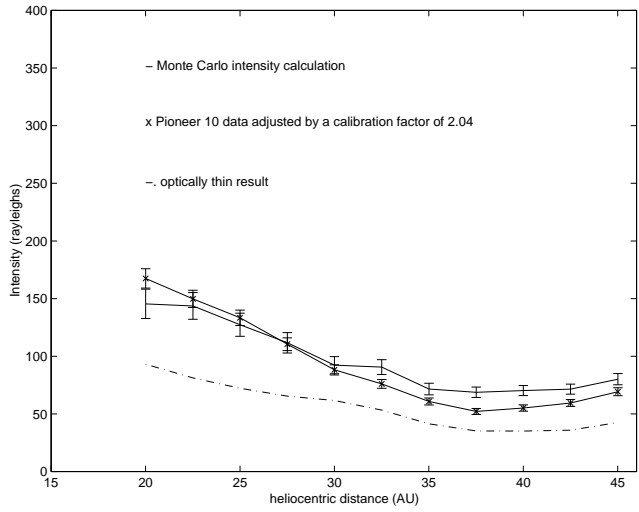}
\caption{ Comparison of Monte Carlo optically thick and optically
thin calculations using heliospheric model with neutral hydrogen
density of 0.15 cm$^{-3}$ and proton density of 0.07 $^{-3}$ and
Pioneer 10 Lyman Alpha glow data adjusted by the constant
calibration factor as function of heliocentric distance. }
\end{figure}
\begin{figure} \label{fig5}
\noindent\includegraphics[width=\hsize]{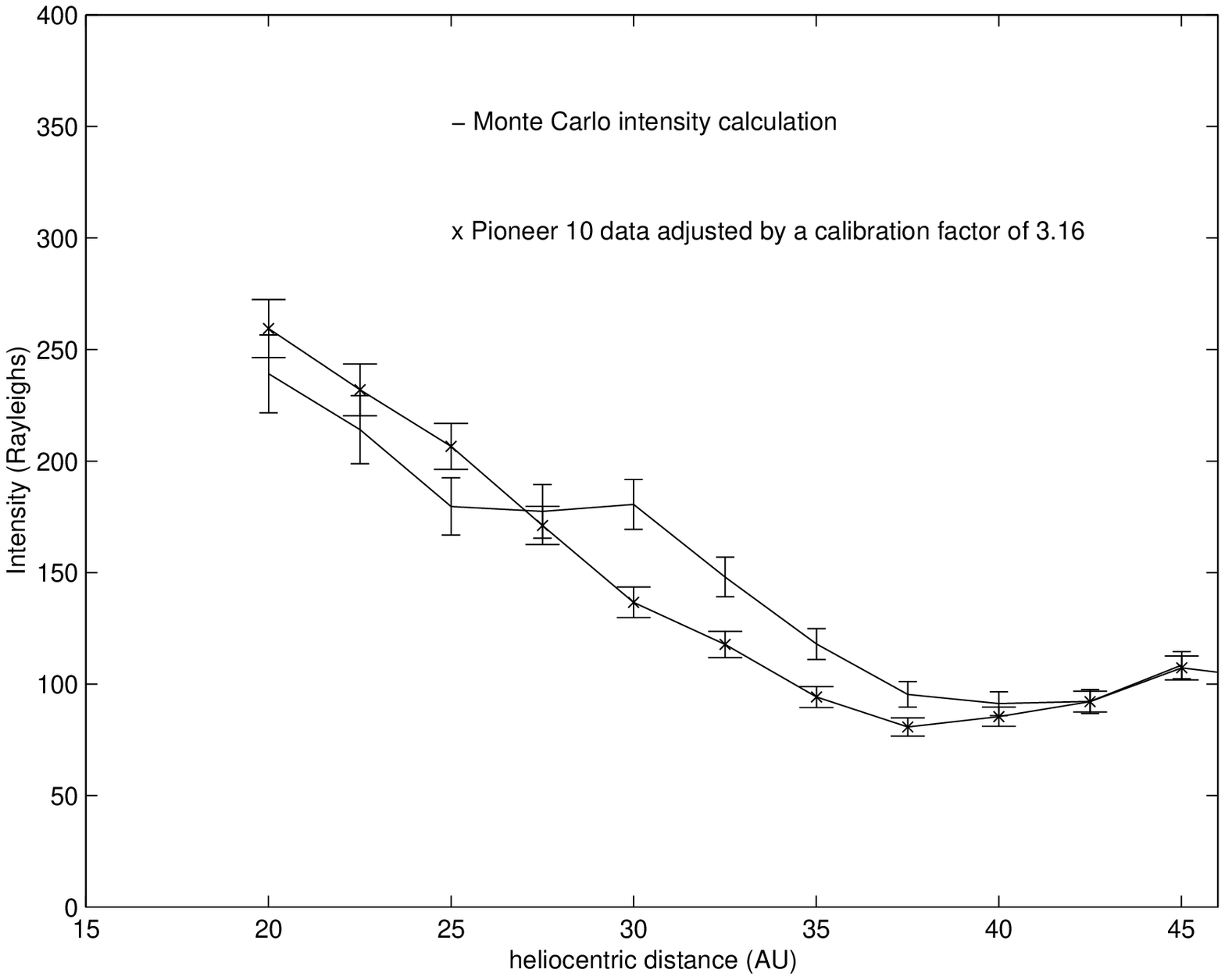}
\caption{Same as figure 4 except that a heliospheric model with
neutral hydrogen density of 0.2 cm$^{-3}$ and proton density of
0.05 cm$^{-3}$ was used}
\end{figure}
\begin{figure} \label{fig6}
\noindent\includegraphics[width=\hsize]{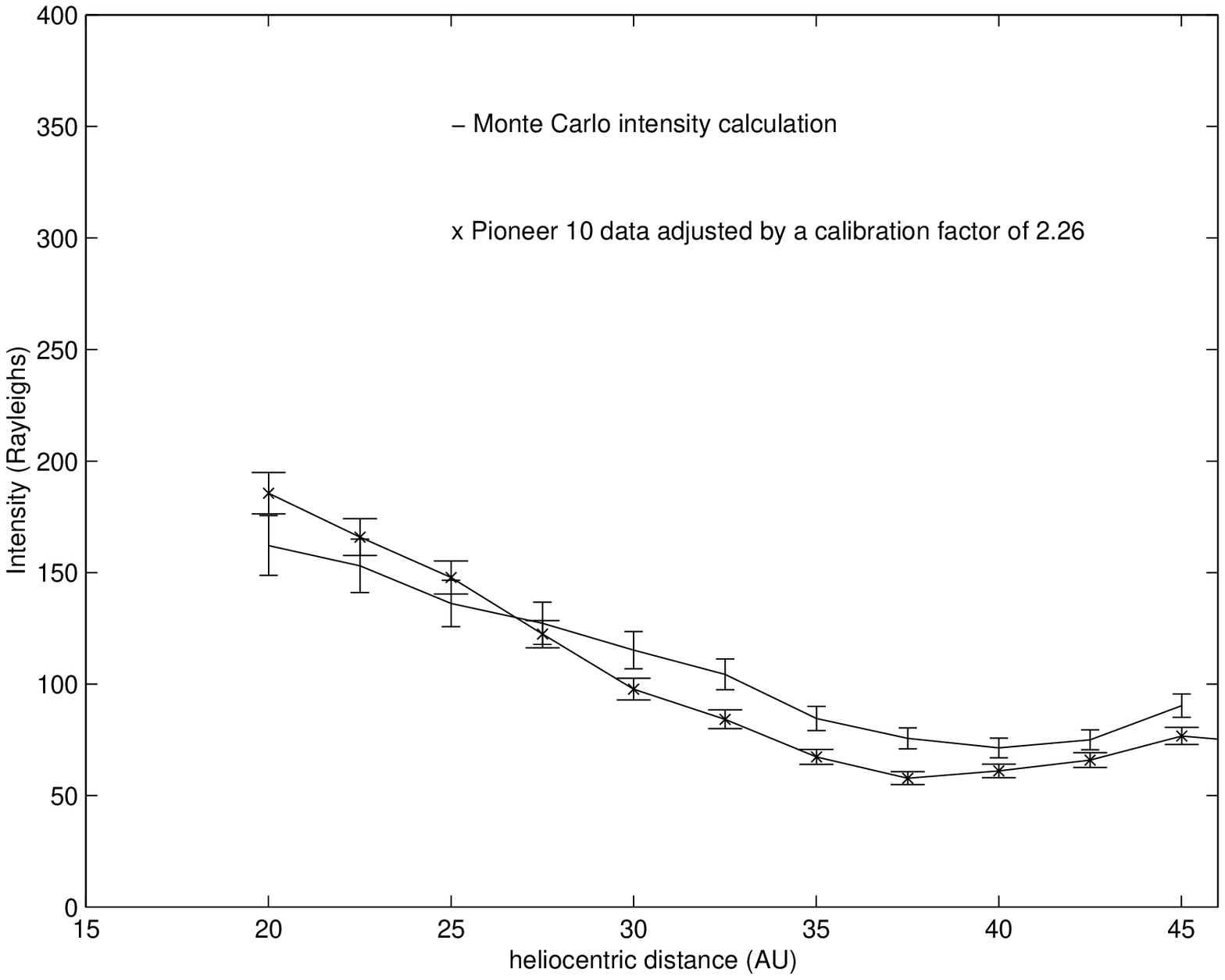}
\caption{Same as in figure 4 except that a heliospheric model with
neutral hydrogen density of 0.2 cm$^{-3}$ and proton density of
0.1 cm$^{-3}$ was used}
\end{figure}
\begin{figure} \label{fig7}
\noindent\includegraphics[width=\hsize]{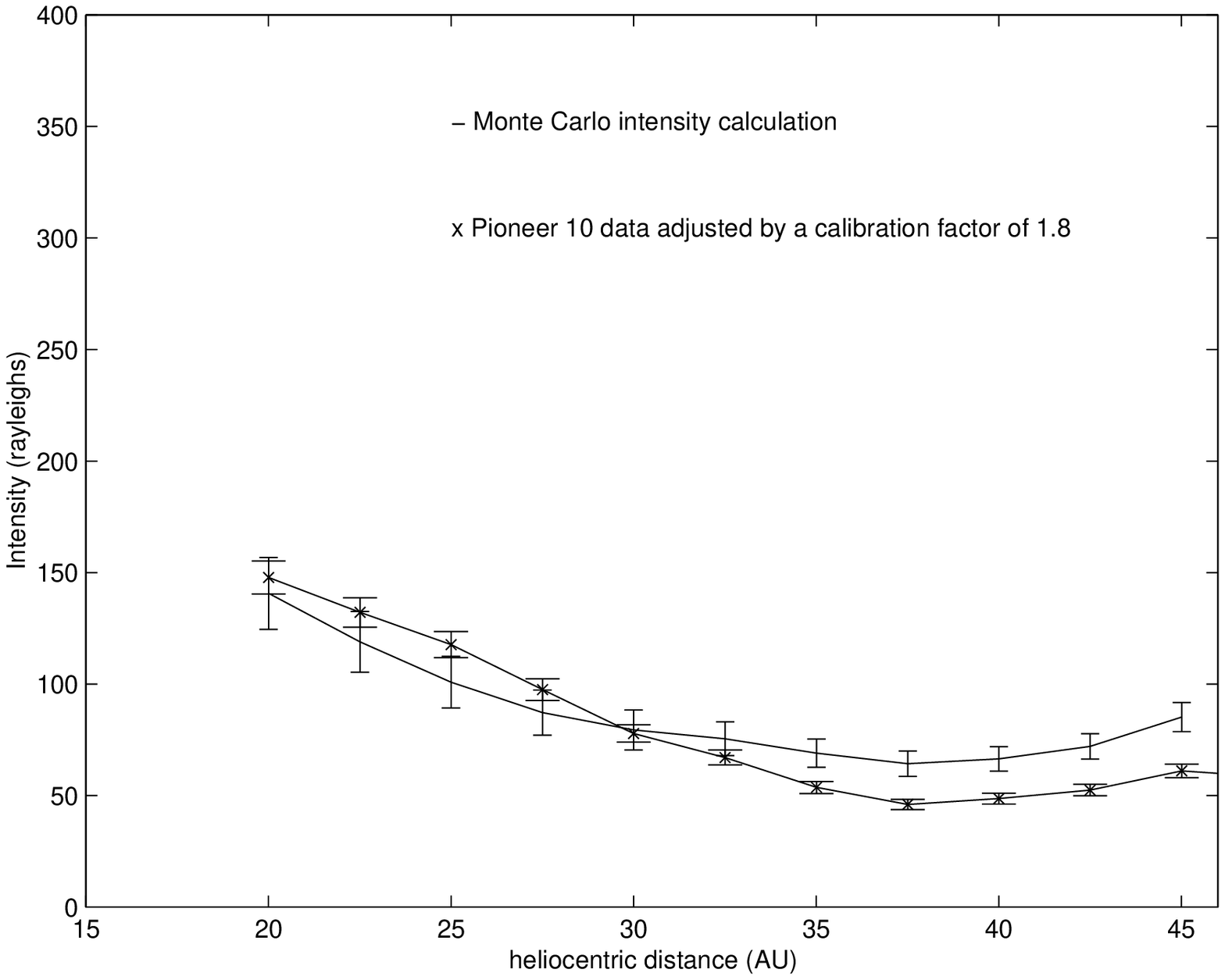}
\caption{Same as in figure 4 except that a heliospheric model with
neutral hydrogen density of 0.2 cm$^{-3}$ and proton density of
0.2 cm$^{-3}$ was used}
\end{figure}

\begin{figure} \label{fig8}
\noindent\includegraphics[width=\hsize]{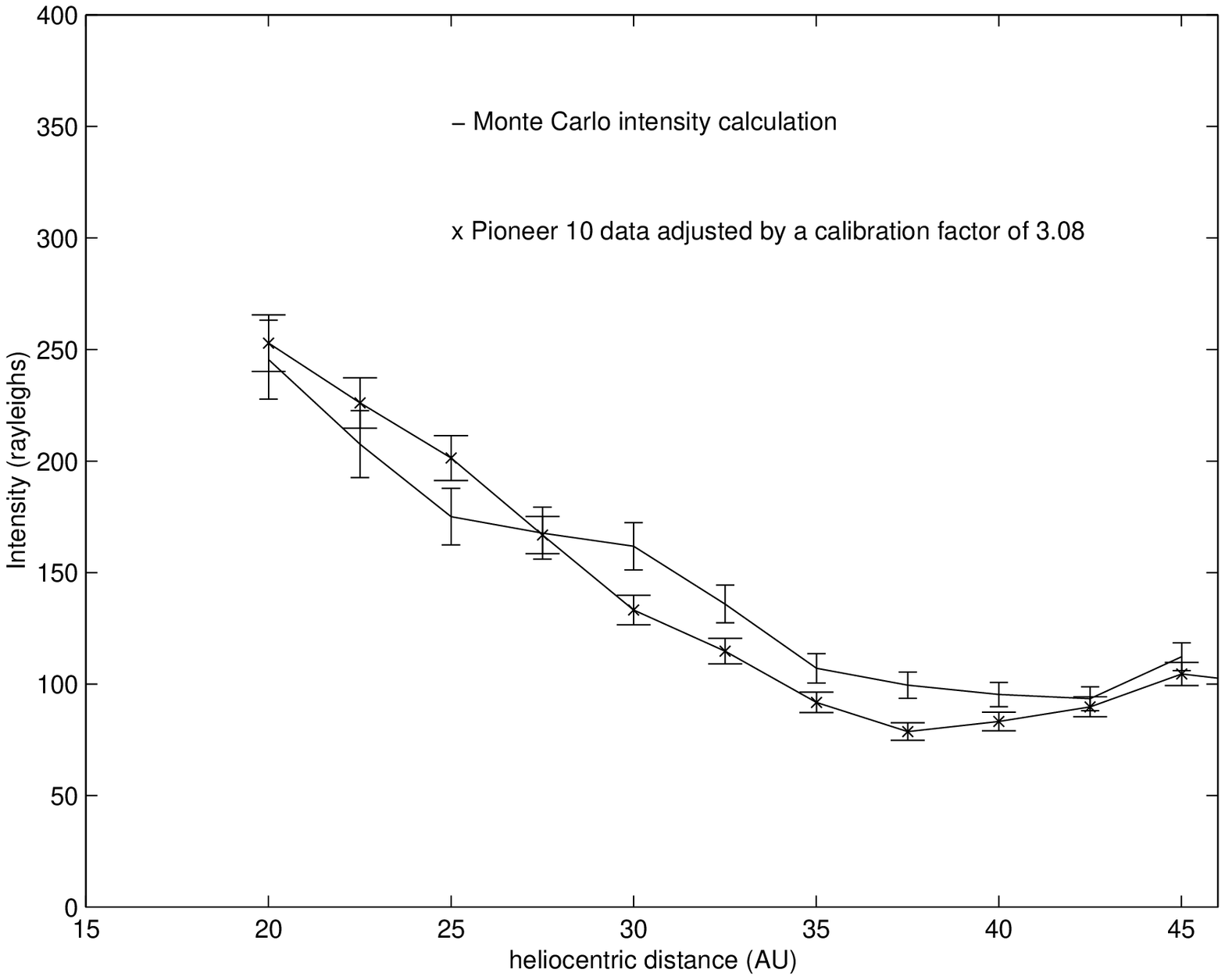}
\caption{Same as in figure 4 except that a heliospheric model with
neutral hydrogen density of 0.25 cm$^{-3}$ and proton density of
0.1 cm$^{-3}$ was used}
\end{figure}
\begin{figure} \label{fig9}
\noindent\includegraphics[width=\hsize]{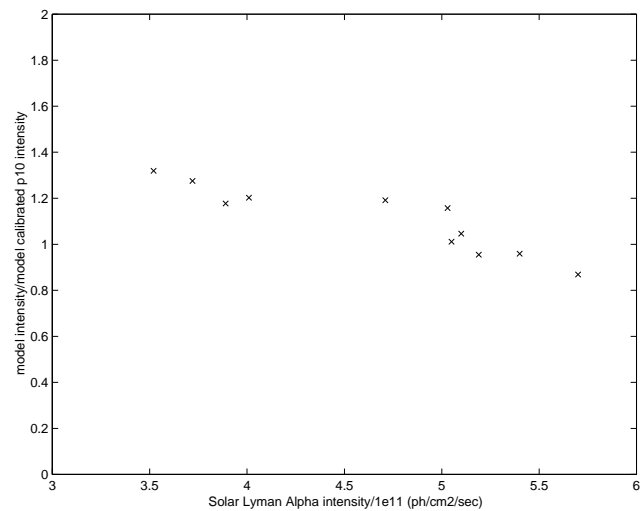}
\caption{The ratio of the model intensity to the model calibrated
P10 data as function of the solar Lyman $\alpha$ flux. The model
intensities are for the best fit heliospheric model with neutral
hydrogen density 0.15 cm$^{-3}$ and proton density 0.07
cm$^{-3}$.}
\end{figure}

\section{Conclusion}

The comparison of predicted Lyman $\alpha$ glow using state of the
art heliosphere  model and Monte Carlo radiative transfer
calculations with P10 data has yielded several constraints on the
VLISM parameters. The ionization ratio is found to vary between
0.2 and 0.5. The upper limit to the neutral hydrogen density is
found to be less than 0.25 cm$^{-3}$. The optimum calibration
factor was found to vary from 1.8 to 3.2 for the five models used
in this work. The calibration factor is found to be 2 for the
model (neutral density =0.15 cm$^{-3}$; proton density = 0.07
cm$^{-3}$) that best fit the data. This work suggests that P10 UV
photometer flux data values need to be increased by a factor of 2.
 There
is some evidence that solar line center flux variation will have
to be taken into account in the interpretation of the P10 data.

\acknowledgements This work was supported by NASA grant
NAG5-10989.
 V.I. was also supported in part by CRDF Award RP1-2248, INTAS 2001-0270, INTAS
YSF 00-163,  RFBR grants 01-02-17551, 02-02-06011, 01-01-00759,
and International Space Science Institute in Bern.

\end{article}

\begin{thebibliography}{}
\bibitem[Ajello et al. (1987)]{Ajello_1987} Ajello, J.M., Stewart, A. I.,
Thomas, G. E., Graps, A., Solar cycle study of interplanetary
Lyman $\alpha$ variations - Pioneer Venus Orbiter sky background
results, Astrophysical Journal, 1, 317, 964-986. 1987.
\bibitem[Axford (1972)]{} Axford, W.I., The interaction of the solar wind
with the interstellar medium, in Solar Wind, NASA Spec. Publ.,
NASA SP-308, 609, 1972.
\bibitem[Baranov et al. (1991)]{blm91} Baranov, V.~B., Lebedev, M.~G., Malama, .~G.,
The influence of the interface between the heliosphere and the
local interstellar medium on the penetration of the H atoms to the
solar system, \apj 375, 347-351, 1991.
\bibitem[Baranov and Malama (1993)]{bm93} Baranov, V.~B., Malama, Y.~G.,
Model of the solar wind interaction with the local interstellar
medium - Numerical solution of self-consistent problem, {\it J.
Geophys. Res.} 98, pp. 15,157-15,163, 1993.
\bibitem[Baranov and Malama (1995)]{bm95} Baranov, V.~B., Malama, Y.~G.,
Effect of local interstellar medium hydrogen fractional ionization
on the distant solar wind and interface region,{\it J. Geophys.
Res.} 100, pp. 14,755-14,762, 1995.
\bibitem[Baranov and Malama (1996)]{bm96} Baranov, V.~B., Malama, Y.~G.,
Axisymmetric Self-Consistent Model of the Solar Wind Interaction
with the Lism: Basic Results and Possible Ways of Development {\it
Space Sci. Rev.} 78,Issue 1/2, p. 305-316, 1996.
\bibitem[Baranov et al. (1998)]{bim98} Baranov, V.~B., Izmodenov, V.~V.,  Y.~G.,
On the distribution function of H atoms in the problem of the
solar wind interaction with the local interstellar medium, {\it J.
Geophys. Res.} 103, 9575-9586,1998.
\bibitem[Cummings et al., 1993]{} Cummings, A.C., Stone, E.C., and Webber, Estimate of the distance of the solar wind termination shock from gradients of
anomalous cosmic ray oxygen, J. Geophys. Res., 98, 15,165, 1993.
\bibitem[Gangopadhyay et al., 1989]{pradip_89}Gangopadhyay, P., Ogawa, H.~S., Judge, D.~L.,
Evidence of a nearby solar wind shock as obtained from distant
Pioneer 10 ultraviolet glow data, \apj, 336, 1012-1021, 1989.
\bibitem[Gangopadhyay and Judge (1995)]{pradip_95}Gangopadhyay, P. and Judge,
D.L., UV remote detection of the solar wind termination shock,
{\it Advances in Space Research}, 15, 8/9, 463-466, 1995.
\bibitem[Gangopadhyay and Judge (1996)]{pradip_96}Gangopadhyay, P. and Judge,
D.~L., Model-insensitive and Calibration-independent Method for
Determination of the Downstream Neutral Hydrogen Density Through
Lyman $\alpha$ Glow Observations, \apj, 467, 865-869, 1996.
\bibitem[Gloeckler et al. (1997)]{gloeckler97}
Gloeckler, G., Geiss, J., Fisk, L., Anomalously small magnetic
field in the local interstellar cloud, {\it Nature} 386, 374-377,
1997.
\bibitem[Gloeckler and Geiss, 2001]{}Gloeckler, G. and Geiss, J., Heliospheric
and Interstellar phenomena deduced from pickup ion observations,
Space Science Reviews, 97, 1/4, 169, 2001.
\bibitem[Hall et al., 1993]{}Hall, D.T., Shemansky, D.E., Judge, D.L.,
Gangopadhyay, P., Gruntman, M.A., Heliospheric hydrogen beyond 15
AU: evidence for a termination shock, {\it J. Geophys. Res.}, 98,
15,185 - 15, 192, 1993.
\bibitem[Holzer, 1972]{} Holzer, T.E., Interaction of the solar wind with the
neutral component of the interstellar gas, J. Geophys. Res., 77,
5407, 1972.
\bibitem[Izmodenov (2000)]{izmod00} Izmodenov, V.~V.,
Physics and Gasdynamics of the Heliospheric Interface, {\it
Astrophys. Space Sci.} 274, 55-69, 2000.
\bibitem[Izmodenov et al., 1999]{izmod99} Izmodenov, V.~V., Lallement, R., a, Yu.~G.,
Heliospheric and it astrospheric hydrogen absorption towards
Sirius: no need for interstellar hot gas, A\&A 342, L13-L16, 1999.
\bibitem[Izmodenov et al., 2001]{izmod01} Izmodenov, V.~V., Gruntman, M., , Yu.~G.,
Interstellar hydrogen atom distribution function in the outer
heliosphere, {\it J. Geophys. Res.} 106, 10681-10690, 2001.
\bibitem[Keller, 1981]{} Keller, H.U., Richter, K., and Thomas, G.E., Multiple
scattering of solar resonance radiation in the nearby interstellar
medium II, Astron Astrophys, 102,415, 1981.
\bibitem[Kurth and Gurnett, 1993]{} Kurth, W.S. and Gurnett, D.A.,
Plasma waves as indicators of the termination shock, J. Geophys.
Res., 98, 9, 1993.
\bibitem[Lallement et al., 1993]{} Lallement, R., Bertaux, J.L., and Clarke, .,
Deceleration of Interstellar Hydrogen at the heliospheric
interface, Science, 260, 1095, May 21, 1993.
\bibitem[Lemaire et al., 1978]{} Lemaire, P., Charra, J., Jouchoux, A.,
Vidal-Madjar, A., Artzner, G.E., Vial, J.C., Bonnet, R.M.,
Skumanich, A., Calibrated full disk solar HI Lyman $\alpha$ and
Lyman $beta$ profiles, \apj, 223, L55, 1978.
\bibitem[Lemaire et al., 1998]{} Lemaire, P., Emerich, C., Curdt, W., Schuhle, ., and Wilhelm, K., Solar HI Lyman $\alpha$ full disk profile obtained with OHO spectrometer, Astron Astrophys, 334, 1095, 1998.
\bibitem[Linsky and Wood, 1996]{} Linsky, J.L., Wood, B.E., The A lpha Centauri
line of sight: D/H ratio, physical properties of local
interstellar gas, and measurement of heated hydrogen (``The
Hydrogen Wall'') near the heliopause, The Astrophys. J, 463, 254,
1996.
\bibitem[Malama (1991)]{malama91}Malama, Y.~G.,Monte-Carlo simulation of  atoms trajectories in the solar system,
{\it Astrophys. Space Sci.} 176, 21-46, 1991.
\bibitem[Mihalas, 1978]{}Mihalas, D., Stellar Atmospheres, W.H. Freeman and company, 1978.
\bibitem[M\"{u}ller et al (2000)] M\"{u}ller, H.~R., Zank,
G.~P., Lipatov, A.~S., Self-consistent hybrid simulations of the
interaction of the heliosphere with the local interstellar medium,
\jgr 106, 27,419 - 27,438, 2000.
\bibitem[Qu\grave{e}merais et al. (1994)]{} Qu\'emerais, E., Bertaux,
J.-L., Sandel, B., Lallement, R., A new measurement of the
interplanetary hydrogen density with ALAE/ATLAS 1,{\it Astronomy
and Astrophysics}, 290, 941-955, 1994.
\bibitem[QuÓmerais et al. (1995)]{eric_95}Qu\'emerais, E., Sandel, B. R.,
Lallement, R., Bertaux, J.-L., A new source of Ly $\alpha$
emission detected by Voyager UVS: heliospheric or galactic origin,
{\it Astronomy and Astrophysics}, 299, 249-257, 1995.
\bibitem[QuÓmerais et al. (1996)]{eric_96}Qu\'emerais, E.,Malama, Y. G.,
Sandel, W. R., Lallement, R., Bertaux, J.-L., Baranov, V. B.,
Outer heliosphere Lyman $\alpha$ background derived from two-shock
model hydrogen distributions: application to the Voyager UVS data,
{\it Astronomy and Astrophysics}, 308, 279-289, 1996.
\bibitem[Scherer and Fahr, 1996]{} Scherer, H., and Fahr, H.J., Lyman
$\alpha$ transport in the heliosphere based on an expansion into
scattering hierarchies, Astron. Asyrophys., 309, 957, 1996.
\bibitem[Scherer and Scherer, 2001]{} Scherer, H., and Scherer,K.,
New results derived from Pioneer 10/11 UV data, in Proceedings of
COSPAR Colloquium on The Outer Heliosphere: The Next Frontiers,
137-149, 2001.
\bibitem[Scherer et al., 1997]{} Scherer, H., Fahr, H.J., and Clarke, J.T.,
Analysis of interplanetary H-Ly Alpha spectra obtained with the  telescope GHRS spectrometer,
 Astron. Astrophys., 325, 745, 1997.
\bibitem[Scherer et al., 1999]{} Scherer, H., Bzowski, M., Fahr, H.J., and Rucinski, D.,
Improved analysis of interplanetary HST-H Lyman Alpha spectra using e-dependent modelings,
Astron. Astrophys., 342, 601, 1999.
\bibitem[Shemansky et al., 1984]{} Shemansky, D. E., Judge, D. L., Jessen, J.
M., Pioneer 10 and Voyager observations of the interstellar medium
in scattered emission of the He 584 A and H Ly$\alpha$ 1216 A
lines In NASA. Goddard Space Flight Center Local Interstellar
Medium, No. 81, 24-27, 1984.
\bibitem[Tobiska et al. (1997)]{}Tobiska, W. K., Pryor, W. R., Ajello, J. M.,
Solar hydrogen Lyman $\alpha$ variation during solar cycles 21 and
22,{\it Geophys . Res. Lett.}, 24, 1123-1127, 1997.
\bibitem[Wang and Richardson (2001)]{} Wang, C., and Richardson, J.D., Energy
Partition between solar wind protons and pickup ions in the
distant heliosphere: A three-fluid approach, J. Geophys.Res. 106,
29401, 2001.
\bibitem[Witte et al. (1993)]{}Witte, M., Rosenbauer, H., Banaszkiewicz, M., r, H., The ULYSSES neutral
gas experiment - Determination of the velocity and temperature of
the interstellar neutral helium, {\it Advances in Space Research},
vol. 13, no. 6, p. 121-130, 1993.
\bibitem[Witte et al. (1996)]{} Witte, M., Banaszkiewicz, M., Rosenbauer, H., cent Results on the
Parameters of the Interstellar Helium from the Ulysses/Gas
Experiment, 1996, {\it Space Science Reviews} 78, Issue 1/2, p.
289-296.
\bibitem[Woods and Rottman (1997)]{} Woods,T.N. and Rottman, G.J., Solar
Lyman $\alpha$ irradiance measurements during two solar cycles,
{\it J. Geophys. Res.}, 102,8769-8779, 1997.
\bibitem[Woods et al. (2000)]{} Woods,T.N., Tobiska, W.K., Rottman, G.J., and
'Worden, J.R., Improved solar Lyman $\alpha$ irradiance modeling
from 1947 through 1999 based on UARS observation, {\it J. Geophys.
Res.}, 105,27,195-27,215, 2000.
\bibitem[Wu et. al. (1981)]{wu81} Wu, F.M., Judge,D.L., Suzuki, K., Carlson,
R.W., Pioneer 10 ultraviolet photometer observations of the
interplanetary glow at heliocentric distances from 2 to 14 AU,
\apj,  245, 1145-1158, 1981.
\bibitem[Wu et. al. (1988)]{wu88} Wu, F.M., Gangopadhyay, P., Ogawa, H.S.,
Judge, D.L., The hydrogen density of the local interstellar medium
ultraviolet observations,  \apj,  331, 1004-1012, 1988.

\end{thebibliography}
\end{document}